# Interplay of phonons, intertwined density waves, and induced spin density wave in trilayer nickelates $Pr_{4-x}La_xNi_3O_{10}$

Sonia Deswal[1, †], Dibyata Rout[2], Nirmalya Jana[3], Koushik Pal[3], Surjeet Singh[2] and Pradeep Kumar[1, *]

[1] *School of Physical Sciences, Indian Institute of Technology Mandi, Mandi-175005, India*
[2] *Department of Physics, Indian Institute of Science Education and Research Pune, Pune-411008, India*
[3] *Department of Physics, Indian Institute of Technology Kanpur, Kanpur-208016, India*

**Abstract**

Lattice degrees of freedom (DoF) play a central role in correlated electron systems, strongly influencing the dynamics of the underlying charge carriers and spin excitations. In nickelates, understanding the role of lattice is essential to unravel the interplay between charge, orbital, and spin degree of freedom in giving rise to various emergent phenomena reported recently. Here, we investigate the phononic DoF in a series of trilayer nickelates, namely $Pr_{4-x}La_xNi_3O_{10}$ (where x = 0, 0.4, 1, 2, 3.6, and 4) using temperature and polarization dependent Raman scattering measurements. Our in-depth analysis of the phonon evolution with temperature and doping, gives interesting insights into the behaviour of these materials. All these systems undergo a metal-to-metal transition ($T_{MMT}$), characterized by the development of intertwined spin and charge density waves, with the spin density wave preceding the charge density wave. These transitions manifest as pronounced anomalies in phonon self-energy parameters i.e. peak frequency and linewidth in the vicinity of the metal-to-metal transition. Several phonon modes show dramatic change (nearly an order of magnitude for some modes) in their softening rates across the $T_{MMT}$, highlighting the sensitivity of the lattice dynamics to spin and charge order. These findings emphasize the crucial role of lattice DoF in mediating correlated ground states in layered nickelates.

**2 December 2025**

†soniadeswal255@gmail.com

* pkumar@iitmandi.ac.in

1. Introduction

Strongly correlated $3d$ and $4d$ transition metal oxides exhibit intriguing magnetic and electronic properties due to their metal ions adopting unusual oxidation states. The emergence of stripe order involving charge- and spin-density wave (CDW and SDW) in these systems has attracted significant interest in condensed-matter community [1–3]. Charge and spin order have been identified in underdoped cuprate superconductors, and have also been observed in non-cuprate transition metal oxides with mixed valence states, such as those containing manganites ($Mn^{3+}/Mn^{4+}$) and nickelates ($Ni^{1+}/N^{2+}$, $Ni^{2+}/Ni^{3+}$) [4–7]. Understanding the interplay of these intertwined order parameters linked with density waves, CDW and SDW, is considered as a frontier challenge in quantum materials such as cuprates and nickelates. The Ruddlesden-Popper (RP) series of the nickelates $R_{n+1}Ni_nO_{3n+1}$ (where R- rare earth metals, La, Pr, and Nd, and alkaline metals such as Sr and Ba), which consists of n $RNiO_3$ perovskite layers, sandwiched between two adjacent rock salt structure type RO layers, have sparked considerable interest due to their structural similarity to cuprates and their potential for high temperature superconductivity [8,9]. The $n = 3$ members of the Ruddlesden-Popper (RP) nickelate series exhibit unusual phase transitions, specifically metal-to-metal transition (MMT), which is reminiscent of the complex metal-insulator transitions observed in cuprates. These transitions are driven by strong electron correlations and are often associated with pseudogap formation [10–12].

The exchange interaction between the $3d$-orbital states of Ni ions and $4f$ electrons of rare earth ion ($R^{+3}$) in the RP series of the nickelates can give rise to a variety of emergent phenomena. Imprinting of $3d$ magnetic ordering onto $4f$ sublattices occurs when the rare earth element itself has intrinsic magnetic moments or its excited states couple with the 3$d$ moments; though in rare cases, the 4$f$ magnetic order may also influence the 3$d$ system [13,14]. The singular property of $R_{n+1}Ni_nO_{3n+1}$ (n=3, R= Pr, La) is the transition from a high-temperature metallic phase to a low-temperature unstable metallic phase dubbed as MMT transition, which has been attributed to the formation of an intertwined charge and spin density wave phase [15–17]. One of the key challenges in nickelates is understanding and controlling how charge and spin influence their properties. In particular the instabilities in the electric band structure is the origin of intertwined spin and charge density waves, which emerges around the onset of MMT [11,15–20]. The MMT in nickelates is believed to be driven by the combined effect of intertwined spin and charge density waves on the Ni sublattice [16,21,22]. Some recent extensive theoretical studies on trilayer nickelates indicate that the SDW primary drives the CDW [17,23,24].

To gain further insight into the formation of concomitant SDW and CDW in these RP nickelates, one should look closely at the structure of these materials. The crystal structure of $R_{n+1}Ni_nO_{3n+1}$ (n=3) nickelates contains two distinct blocks: one corresponding to the rocksalt layer (RS), and the other being the perovskite block layer (PB). In the specific case of $Pr_4Ni_3O_{10}$, the RS layer consists of ninefold coordinated $Pr^{3+}$ ions that exhibit a crystal-field doublet ground state at low temperatures; while $Pr^{3+}$ ions in the PB layer, are in a twelve-fold coordination, and have a crystal-field split nonmagnetic singlet ground state, a comparable scenario has been observed for the Pr ions in, for example, $Pr_3RuO_7$ and $PrNiO_3$ [25,26]. $Pr_4Ni_3O_{10}$ exhibits density wave transitions on Ni and Pr sublattices at ~ 158 K and 5 K, respectively. Interestingly, the nature of intertwined density waves on the Ni sublattice is quasi-two-dimensional (2D). It is advocated that with lowering temperature this quasi 2D SDW transforms into long-range ordered 3D structure of stacked SDWs at ~ 26 K [17]. Additionally, it induces a 3D SDW on the Pr sublattice around this temperature (~26 K), which would otherwise be nonmagnetic due to the singlet nature of the non-Kramers $Pr^{3+}$ ion in a low-symmetry site.

This SDW transition on Pr sublattice is in addition to the long-range ordered state reported at ~ 5 K [20]. As the concentration of La at Pr sites increases, $La^{3+}$ being non-magnetic in nature, the possibility of induced SDW as well as long-range ordering of the rare earth ions in the RS layer diminishes at low temperatures. The dynamics of phonons are strongly affected by the underlying changes in the spin/charge structure and Fermi-surface reconstruction owing to opening of the spin/charge gapped state. The same may be captured using an in-depth study of the evolution of phonons and underlying quasi-particle excitations. It is also conjectured that superconducting pairing in these systems is induced due to partial nesting between the M=($\pi$, $\pi$) centered pockets and portions of the Fermi-surface centered at the $\Gamma$=(0, 0) point [27]. Therefore, the renormalization of the phonon modes near the Gamma point implicates an intricate link with the pairing mechanism, pointing towards the crucial role of phonons in the pairing process.

Interestingly, spectroscopic studies on $Pr_4Ni_3O_{10}$ that explore phonon dynamics as a function of temperature across the transition into density wave phases are still lacking. In underdoped cuprate superconductors and other strongly correlated metallic systems, enhanced electronic correlations often lead to phonon softening [28]. In fact Raman scattering has proven to be a powerful technique for investigating dynamic phonon coupling associated with complex, temperature-dependent phase transitions in a wide range of correlated materials [29–34]. In particular, anomalies in the phonon spectra and modifications in the electronic background have been widely reported as signatures of CDW formation [32,35]. Similarly, Raman studies have provided evidence of SDW formation in iron-based compounds [34].

Here, we report an in-depth temperature-dependent Raman study to decipher the impact of rare-earth site substitution on the MMT transition, evolution of CDW, SDW and their intertwining, and evolution of the phonon dynamics in $Pr_{4-x}La_xNi_3O_{10}$ (x = 0, 0.4, 1, 3.6, and 4). We observed that MMT transition temperature $T_{MMT}$, depends strongly on the radius of $R^{+3}$ ion: As one moves from Pr to La, the cell parameter increases, and the $T_{MMT}$ changes from ~ 156 K (Pr) to ~ 136 K (La), evidenced by our specific heat, resistivity, and Raman measurements. Here, we demonstrate that the high temperature metal-to-metal transition around $T_{MMT}$ results from the combined effect of spin and charge density waves on the Ni sublattice, with the SDW likely preceding the CDW; as evidenced by the strong anomalous renormalization of the phonon modes. Additionally, our measurements revealed the presence of a Ni sublattice-induced SDW wave on the Pr sublattice, which persists well above the temperature of long-range spin-order on the Pr sites. This SDW systematically diminishes with increasing La concentration. Furthermore, we also observed strong coupling of the crystal field excitations and low-frequency phonon modes, particularly in samples with high Pr-concentration (i.e. for x ≤ 1).

## 2. Experimental and computational details

**2.1 Experimental details**

The polycrystalline samples of $Pr_{4-x}La_xNi_3O_{10}$ with x = 0, 0.4, 1, 2, 3.6, and 4 were synthesized using a wet chemical method. This approach was chosen to ensure the formation of pure-phase compounds, thereby avoiding the mixed phases and intergrowths that can occur with conventional solid-state synthesis methods. For further details of sample preparation and characterization, see Ref. [20]. All samples were characterized by lab-based powder X-ray diffraction (XRD) at room temperature. To determine the crystal structure and phase purity, XRD profile refinement was performed by using the Rietveld method with the FULLPROF suite software, and the pseudo-Voigt function was used to model the line profile [37]. The resistivity and specific heat measurements were carried out using a Physical Property Measurement System (PPMS), Quantum Design USA.

Temperature-dependent Raman measurements were performed in backscattering geometry using a Horiba HR Evolution spectrometer equipped with Peltier-cooled charged coupled device detector to collect the scattered light. The sample was excited using a solid-state laser with wavelength of 532 nm, to prevent any impacts of local heating the laser power was kept very low, < 1 mW. A 50× long working distance objective lens was used to focus the laser on the sample surface. We used a large groove density grating (1800 grooves/mm) for the higher spectral resolution, which dispersed the Raman signals onto the detector. A closed cycle He cryostat (Montana Instruments) was used for variation of temperature in the range from 6 K to 330 K, with ~ ± 0.1 K accuracy. To understand the symmetry characteristics of the phonon modes, we performed polarization dependent Raman measurements at 6 K, using half-wave plate as an analyzer for the incident light and a polarizer for the scattered light.

To determine the peak frequency, full width at half maximum (FWHM), and intensity of the phonon modes from the Raman spectra, we employed ORIGIN software to fit the data with a series of Lorentzian functions. The fitting function used is, $Y(\omega) = Y_0(\omega) + \frac{2A}{\pi} \frac{\Gamma}{4(\omega - \omega_C)^2 + \Gamma^2}$, where A, $\Gamma$, and $\omega_c$ are the intensity, FWHM, and peak frequency of a particular phonon mode, respectively. After performing baseline correction and selecting the appropriate number of peaks to fit, we iterated the process until the best possible fit to the spectrum was achieved.

**2.2. Computational details**

We used Vienna ab initio simulation package (VASP) [38,39] for the density functional theory (DFT) calculation within a plane-wave basis set. The generalised gradient approximation (GGA) of type Perdew-Burke-Ernzerhof (PBE) [40] exchange-correlation and projector augmented wave potentials [39,41] are used for the constituent elements with the valence configurations La ($5s^2 5p^6 5d^1 6s^2$), Ni ($3d^9 4s^1$) and O ($2s^2 2p^4$). We used DFT+U method to treat the partially filled d-orbitals. The effective Hubbard interactions $U_{eff}$ [42] for transition metal Ni atoms with correlated 3d orbitals are treated in the Dudarev scheme [43]. The $U_{eff}$ for Ni is chosen to be 3 eV [44]. We set the kinetic energy cutoff of the plane wave basis as 520 eV. We do the symmetry-protected volume and ionic relaxation by the conjugate-gradient algorithm until the Hellman Feynman forces on each atom reach a value below the tolerance value of $10^{-4}$ eV/°A. We calculated the harmonic phonon modes at the gamma point of $La_4Ni_3O_{10}$ using the density functional perturbation theory as implemented in the VASP [45] code.

## 3. Results and discussion

The room temperature XRD pattern of $Pr_4Ni_3O_{10}$ is shown in Fig. 1(a). All the observed Bragg peaks are well indexed, and the refined XRD data confirm that the sample is formed in a pure phase with the space group $P2_1/a$. The elementary cell of $Pr_4Ni_3O_{10}$ (monoclinic), contains Z=4 formula units. The structure exhibits anomalous changes in the lattice parameters across the MMT ~ 158 K, without any change the crystal symmetry [20]. We confirm homogeneous substitution of La at the Pr site using XRD data, and investigate the effect of doping on lattice parameters. The room temperature XRD data for other samples $Pr_{4-x}La_xNi_3O_{10}$ (x = 0.4, 1, 2, 3.6 and 4) are shown in Supplementary Information [46] (Fig. S1). We observed a monotonous shift of the XRD peaks towards lower 2θ without any additional splitting or the appearance of new peaks, indicating that the phase remains consistent across all La doping concentrations. Figure 1(b) shows the extracted lattice parameters *a*,*b*, and *c* as a function of doping concentration. The lattice parameters shows an increasing trend with increasing La doping, due to the larger ionic radius of $La^{3+}$ compared to that of $Pr^{3+}$.

In Fig. 1(c), we plot the specific heat ($C_P$) as a function of temperature for all the compositions. A sharp anomaly is observed for all the compositions at their respective metal-to-metal transitions temperatures. For x = 0, the anomaly is observed at ~156 K, and as the La-doping levels increases, the anomaly shifts to lower temperatures to ~ 136 K for x=4, $La_4Ni_3O_{10}$. The low-temperature $C_P/T$ exhibits a broad peak for x = 0, 0.4, and 1.0 in addition to the lambda shaped anomaly seen at the MMT. The inset of Fig. 1(c) shows the zoomed-in region around low temperatures to highlight this anomaly. For $Pr_4Ni_3O_{10}$, the low-temperature anomaly is found to be at $T_1$ ~ 4.8 K and suppresses significantly with increasing La substitution at Pr site. The low-temperature anomaly at $T_1$ is attributed to the magnetic ordering of $Pr^{3+}$ ions in the RS layers, which exhibits a ground state doublet with an antiferromagnetic ordering below $T_1$. $Pr^{3+}$ exhibits partially filled $f$-shells ($4f^2$), which makes magnetic ordering possible for $Pr_4Ni_3O_{10}$. In contrast, $La^{3+}$ has a completely filled electronic configuration for the 4$f$ orbital, which makes it nonmagnetic. As the La concentration at the Pr site increases, this peak becomes progressively suppressed and completely disappears at a doping level of 50%. This phenomenon may be attributed to the reduction of antiferromagnetic ordering caused by doping with non-magnetic ions, leading to a decrease in the magnetic contribution to the total specific heat. We also observe a broad peak in the magnetic heat capacity at $T$~36 K, which is due to the Schottky anomaly occurring from crystal field splitting of the lowest J=4 multiplet of the $Pr^{3+}$ ion in PB layers of $Pr_4Ni_3O_{10}$ [20]. It is observed that the Schottky anomaly decreases with increasing La doping at the Pr site. This can be attributed to the reduction in the concentration of magnetic $Pr^{3+}$ ions and the corresponding increase in non-magnetic $La^{3+}$ ions. Since $La^{3+}$ lacks 4f electrons, it reduces the number of discrete energy levels responsible for the Schottky contribution. As a result, the amplitude of the Schottky anomaly gradually diminishes and eventually disappears at high La concentrations.

Figure 1(d) shows the evolution of MMT transition ($T_{MMT}$) as a function of La concentration extracted from specific heat, resistivity, and Raman measurements (see Supplementary Information [46] Fig. S2 (b) for resistivity measurements). It is found that $T_{MMT}$ systematically decreases with increasing La concentration, suggesting that MMT transition is significantly dependent on rare earth ionic radius. The MMT transition temperature extracted from both specific heat and resistivity are closely aligned to those extracted from anomalies in the phonon mode's self-energy parameters, as discussed in detail in sections 3.1 and 3.2.

Group-theoretical analysis predicts [47] $\Gamma_{Total} = 48A_g + 54A_u + 48B_g + 54B_u$ phonon modes, out of which $A_g$ and $B_g$ are Raman active modes (see Table S1 for more details). Figure 2 (a) shows the Raman spectrum of the undoped sample Pr$_4$Ni$_3$O$_{10}$ (x = 0) in the spectral range of 30-680 cm$^{-1}$ recorded at 6 K. The spectra are fitted using a sum of Lorentzian functions to extract the self-energy parameters of the phonon modes, i.e., mode frequency ($\omega$) and full width at half maximum (FWHM)/ linewidth, as well as the intensity. Fourteen distinct Raman modes at 6 K, which, for convenience are labeled P1-P14. Their tentative $A_g$ symmetry were identified from polarization analysis, as shown in Fig. S3 (see Supplementary Information [46] section S1 for more details). Based on earlier neutron scattering studies [20,26], an additional mode C1 (~117cm$^{-1}$) observed at low temperatures up to ~ 150 K, is attributed to the crystal field excitation associated with the Pr$^{3+}$ion ($^3$H$_4$, S=1, L=5, J=4). Figure 2(b) shows temperature evolution of the Raman spectrum of Pr$_4$Ni$_3$O$_{10}$ in the range 6-330 K. With increasing temperature, the peaks become broader, and some of the red bar highlighted peaks gradually disappear. For the La$_4$Ni$_3$O$_{10}$, we obtained 33 distinct modes, labeled S1-S33. A comparison of phonon frequencies obtained from experimental data and first-principles DFT calculations is provided in Supplementary Information, Table S2 [46].

## 3.1. Temperature dependence of the phonon self-energies in Pr$_4$Ni$_3$O$_{10}$

Figure 3 illustrates the temperature dependence of the phonon self-energy parameters i.e. peak frequency $(\omega)$ and full-width at half maxima (FWHM) or linewidth for several prominent phonon modes in Pr$_4$Ni$_3$O$_{10}$. The following key observations emerge: (i) Anomalies near the vicinity of MMT temperature $T_{MMT}$ (~ 160K). Most phonon modes exhibit anomalous renormalization in both frequency and linewidth near T$_{MMT}$. All phonon modes except P2 display typical anharmonic behavior, with increasing peak frequency upon cooling from 330 K down to T$_{MMT}$. Modes P1, P6, P9 (not shown), P10, and P14 show a distinct change in slope around T$_{MMT}$. Notably, below T$_{MMT}$, the softening rate of modes P1, P6, and P10 decreases markedly i.e., $\Delta\omega(\omega_{T_{MMT}} - \omega_{6K}) \sim 1$ cm$^{-1}$, while above T$_{MMT}$, these modes exhibit a pronounced increase in softening rate (i.e., $\Delta\omega(\omega_{330K} - \omega_{T_{MMT}}) \sim 2$ to $5$ cm$^{-1}$). Mode P14 displays a lower softening rate below $T_{MMT}$ i.e., $\Delta\omega(\omega_{T_{MMT}} - \omega_{6K}) \sim 1$ cm$^{-1}$, while, it shows the most significant softening, deviating by $\Delta\omega(\omega_{330K} - \omega_{T_{MMT}}) \sim 9$ cm$^{-1}$ below $T_{MMT}$. (ii) Frequency of the mode P2 (~85 cm$^{-1}$) exhibits anomalous hardening with increasing temperature till ~ 160 K, which is diagonally opposite to the expected behavior of a normal phonon mode and it is nearly constant from 330 K to ~ 160 K. This anomaly is understood by invoking coupling of mode P2 (~85 cm$^{-1}$) with the CFE observed around ~117cm$^{-1}$.

Linewidths of all modes (except P10) show normal behavior i.e., linewidth decreases with decreasing temperature from 330 K to $T_{MMT}$. The linewidth of the modes P1, P2, P4, P7, P10, and P14 show a change around $T_{MMT}$. Quite interestingly, with decreasing temperature till 30 K linewidth of the modes P2 and P4 shows broadening below $T_{MMT}$. On the other hand linewidth of the modes P1, P7 and P10 show a dome like structure below $T_{MMT}$ i.e., linewidth increases with a decrease in temperature and reaches a maximum and decreases thereafter down to ~ 30 K. Generally, the onset of a CDW leads to a decrease in phonon linewidth, likely due to instabilities in the electronic structure and reduced electron-phonon coupling, which in turn increases the phonon lifetime and decreases the scattering rate [32,35]. Therefore, the significant increase in the linewidths of modes P2 and P4 below MMT, along with the dome-like behavior observed in a few other modes, suggest the emergence of a SDW below $T_{MMT}$. This likely results from the introduction of additional decay phonon channels associated with spin fluctuations, leading to enhanced linewidths. These anomalies suggest that the SDW precedes and potentially drives the formation of the CDW below $T_{MMT}$. Specifically, the initial increase in linewidth can be attributed to the SDW, while the subsequent decrease at lower temperatures may reflect the dominance of the CDW, which typically narrows linewidths due to reduced scattering. Similar dome-like features are observed in $La_4Ni_3O_{10}$ near the metal-to-metal transition, supporting the existence of intertwined density waves emerging in this class of materials [48].

To understand the deviation in the mode frequency and linewidth around $T_{MMT}$, from their normal behaviour, we have used simple cubic anharmonic phonon-phonon interaction model [49,50]. Within this model, the peak frequency and linewidth may be expressed as:

$$\omega(T) = \omega_0 + A\left(1 + \frac{2}{e^x - 1}\right), \text{ and } \Gamma(T) = \Gamma_0 + B\left(1 + \frac{2}{e^x - 1}\right)$$

where $\omega_0$, and $\Gamma_0$ are the mode frequency and FWHM at absolute zero temperature, respectively and $x = \frac{\hbar\omega_0}{2k_B T}$, and constant parameters A and B are linked to the three-phonon anharmonic decay process. As seen in Fig. 3, the temperature dependence of the prominent phonon modes fits well using anharmonic model within the range of 330 K to ~160 K (solid red curve) (see Fig. S4 in the Supplementary

Information [46] for other modes). We extended this anharmonic curve below $T_{MMT}$ (as shown by the dotted lines), to the lowest measured temperature i.e. 6 K. Phonon modes show pronounced deviation from the expected curve of a simple anharmonic phonon-phonon interaction model below $T_{MMT}$, suggesting that the conventional anharmonic effects cannot account for the anomalous renormalization of phonon self-energies. The observed phonon renormalization below $T_{MMT}$ (~ 160 K) reveals the coexistence of the different ordering phenomena such as CDW and SDW [32,51]. Strong renormalization of the phonon modes below $T_{MMT}$ evidence that nonphononic contribution to phonon self-energies are dominating. In general, the emergence of a CDW leads to phonon mode softening and linewidth narrowing, primarily due to instabilities in the electronic structure, specifically the Fermi surface, lead to these density wave formations [17]. Such behavior has been reported in various material systems [35,52,53]. In particular, the 2H-TaS$_2$ compound exhibits softening of two-phonon features due to short-range CDW order, and CsV$_3$Sb$_5$ shows phonon renormalization around its CDW transition [32,35]. These observations are consistent with the phonon renormalization we observe across the intertwined density state. On the other hand, the emergence of SDW gives rise to linewidth broadening due to existence of additional decay channel resulting from magnons band. Our observation of a dome-like structure of the linewidth for some of the modes (P1, P7 and P10 ), suggesting that SDW sets in before CDW, is also advocated in other reports [17].

Additionally, renormalization of the self-energy parameters of the phonons may be used to capture the influence of short-range ordering of spins, which may begin much above the long-range ordering temperature. It has been reported that below ~ 26 K, Pr$_4$Ni$_3$O$_{10}$ exhibits an induced SDW on the Pr sublattice imprinted by the ordering of the Ni spins [17]. Quite interestingly, some phonon modes exhibit a change in the slope of their frequencies (P2, P6, P10, P11, P14) and FWHM (P1, P2, P4, P6, P7, P10, P11) near ~30 K($T_S$), indicating the onset of induced SDW on the Pr sublattice, see Fig. 3. The changes in the mode frequency around ~ 30 K ($T_S$) are small, however corresponding changes in the linewidth are quite appreciable. To decipher the changes below ~30 K, in particular in the mode frequency, we carried out a temperature-dependent measurement at much smaller temperature interval (~ 2 K) from ~ 6 K to ~ 50 K. Our observation shows change in the temperature variation of mode frequencies around $T_S$, for some of the prominent modes see Fig. 4(a). The anomalous increase of FWHM below ~ 30 K may be understood due to the onset of induced SDW of spin below $T_S$. As the system begins to enter the induced SDW phase below ~30 K, more decay channels will be available for the phonons. The optical phonon modes will have more decay channels, leading to a shorter lifetime $(\tau)$, and an increase in the FWHM $\propto 1/\tau$ below $T_S$, as observed. However, the degree of renormalization differs for different phonon modes, as it is influenced by the mode energy as well as symmetry of the individual phonon. This ordering begins to develop far above the long-range ordering ($T_1$ ~ 4.8 K) of the Pr-4$f$ moments suggesting the existence of SDW below ~ 30 K on the Pr sublattice.

### 3.2. Effect of La doping in Pr$_{4-x}$La$_x$Ni$_3$O$_{10}$

Figure 5(a) shows the evolution of the Raman spectra of Pr$_{4-x}$La$_x$Ni$_3$O$_{10}$ with La doping (x = 0, 0.4, 1, 2, 3.6, and 4) at 6 K. To extract the mode frequency $(\omega)$, FWHM, and intensity of each mode, the spectrum is fitted with a sum of Lorentzian functions, allowing us to distinguish 14(33) modes that have been labeled as P1-P14(S1-S33) for x = 0(x = 4) (For the fitting, labeling and peak positions refer to Fig. S5 and Table S3, respectively in the Supplementary Information [46] ). We note that with increasing La doping, additional weak phonon modes emerge, and the existing modes become sharper compared to those at lower doping levels. In particular, XRD data for La$_4$Ni$_3$O$_{10}$ is refined with a mixture of two phases, namely orthorhombic (*Bmab*) and monoclinic (*P2$_1$/a*) and the corresponding phase fraction is found to be *P2$_1$/a* : *Bmab* = 86.3:13.7 [20]. With increasing La doping, additional weak phonon modes emerge. However, all the prominent phonon modes remain consistent across the full doping series (x = 0, 0.4, 1, 2, 3.6, and 4). When the rare-earth ion is changed from Pr to La, the phonon modes are expected to remain largely the same because the dominant phase is common in both cases, consistent with observations reported for pyrochlore iridates $A_2Ir_2O_7$ (A = Gd, Dy, and Er) [54]. Since Raman signal intensity is approximately proportional to the phase fraction, the dominant phase produces much stronger Raman signals than the minor phase. Therefore, the appearance of weak additional modes in La-rich samples can be attributed to the small fraction of the secondary *Bmab* phase.

Most phonon modes (except mode P14/S32) soften significantly with increasing La concentration. This may be attributed to the differences in ionic radius of Pr/La-cite cation and the different electronic configuration of $Pr^{3+}$ and $La^{3+}$ ions. The phonon modes P4, P10, and P11 exhibit significant softening rates $\Delta\omega \sim 15, 11,$ and $25$, respectively; see Figs. 5(b-e). Within the harmonic approximation framework, the theoretical explanation for the Raman mode frequency $\omega$ involves its connection to the elastic constant and reduced mass $m_r$ of the system, given as: $\omega^2 \propto \frac{k}{m_r}$. Presence of the larger $La^{3+}$ $(1.28 \text{ Å})$ ion as compared to smaller $Pr^{3+}$ $(1.17 \text{ Å})$ ion causes the unit cell parameter and bond lengths to increase, which reduces the elastic constant $k$ and may causes frequency blue shift in phonon modes as observed here for majority of the modes. The reduction in the mass from 140.9 amu (for $Pr^{3+}$) to 138.9 amu (for $La^{3+}$) contributes to the shifts of the Raman peaks to a higher frequency. However, this mass effect is minimal compared to the significant impact of the ionic radius on the phonon mode frequencies (except for mode P14). The inset in Fig. 5(a) depicts the variation of intensity of the P14 (S32) mode as a function of La doping. Frequency of the mode P14 shows an anomalous shift in contrast to other phonon modes, especially for x ≤ 2. Mode P14(S32) directly involves the oxygen stretching vibration along the z-axis around the $Pr^{3+}/La^{3+}$ ions. As P14(S32) mode corresponds to the $La^{3+}$ concentration, reduction of $m_r$ from $La^{3+}$ to $Pr^{3+}$ may leads to redshift below $x \leq 2$. Overall, the substitution of Pr with La alters both phononic and electronic properties, causing shifts in phonon frequencies and line widths due to the coupling between lattice vibrations and the electronic order.

To gain a better understanding of MMT and possible short-range spin-ordering at low temperatures for doped samples, we will now explore temperature-dependent phonons in detail for $Pr_{4-x}La_xNi_3O_{10}$ series at different La doping i.e $x = 0.4, 1, 2,$ and $3.6$ as shown in Fig. 5 (a) see Supplementary Information [46] Fig. S6, S7 for temperature evolution of the Raman spectra across all doping). Following important observations can be made: (i) With decreasing temperature, the phonon modes show deviation from the normal behavior, as expected within the cubic anharmonic phonon-phonon model in the vicinity of $T_{MMT}$. In particular frequencies of the modes P1, P4, P7, P10, P11, and P14 for $Pr_{3.6}La_{0.4}Ni_3O_{10}$ ($x = 0.4$) show deviation from the anharmonic phonon-phonon interaction model below $T_{MMT} \sim 155$ K (see Fig. 6(a)). It is noted that softening rate of the phonon mode P10 and P14 are small below $T_{MMT}$ $\Delta\omega(\omega_{T_{MMT}} - \omega_{6K}) \sim 0.5$ to $1$ cm$^{-1}$, whereas between $T_{MMT}$ to 330 K the corresponding changes are much larger $\Delta\omega(\omega_{330K} - \omega_{T_{MMT}}) \sim 5$ to $10$ cm$^{-1}$. These strong renormalizations of modes P4, P10, and P14 in $Pr_{3.6}La_{0.4}Ni_3O_{10}$ ($x = 0.4$) suggest the emergence of an intertwined density state below $T_{MMT}$, as highlighted by the deviations of these modes in the shaded region in Fig. 6(a). Figure 6(b) shows the phonon modes P3, P4, P5, P9, P11, and P14 of ($x = 1$) $Pr_3La_1Ni_3O_{10}$, where we observe renormalization effects around $\sim 150$ K. The pronounced deviations of the modes P4, P9 and P14 from the normal behavior, indicate the presence of an intertwined density wave. In particular, modes P9 and P14 exhibits weak softening rates below $T_{MMT}$ $\Delta\omega(\omega_{T_{MMT}} - \omega_{6K}) \sim 0.5$ to $2.5$ cm$^{-1}$ but display strong softening rate above $T_{MMT}$ i.e. $\Delta\omega(\omega_{330K} - \omega_{T_{MMT}}) \sim 3$ to $14$ cm$^{-1}$ (ii) For doping values of $x = 0.4$ and $1$ frequency of the mode

P2 remain nearly constant from 330 K to ~ 150 K and shows anomalous softening with further decrease in the temperature till ~ $T_S$ (~ 25 - 20 K) and below ~ $T_S$ it again becomes nearly constant till ~ 6 K. This anomalous behaviour of P2 mode is observed only for higher concentration of Pr i.e. for x = 0, 0.4, and 1 suggesting its strong coupling with the crystal field excitation of $Pr^{3+}$.

Quite interestingly, similar to x = 0 case, $Pr_{3.6}La_{0.4}Ni_3O_{10}$, and $Pr_3La_1Ni_3O_{10}$ shows the phonon renormalization at $T_S$, though the changes are small. We observed a small change in slope of the frequency near $T_S$ ~ 25 K for modes P1, P2, P11, and P14 of (x = 0.4) $Pr_{3.6}La_{0.4}Ni_3O_{10}$ as shown in Fig. 6(a). A similar slope changes in frequency for the modes P3, P7, P9, P11 and P14 is observed for $Pr_3La_1Ni_3O_{10}$ at ~ 20 K (see Fig. 6(b)). To further decipher these small changes near $T_S$, we performed our measurements at much smaller temperature interval for $Pr_3La_1Ni_3O_{10}$ (see Fig. 4(b))). The temperature evolution of the frequency and FWHM for the modes P7, P9, and P11 for $Pr_3La_1Ni_3O_{10}$ and we do observe a change in frequency and FWHM around ~20 K as shown in Fig. 4(b). It is noted that as La doping increases the magnitude of the low temperature anomalies around $T_S$ decreases and eventually disappears at higher La concentrations. This revealing that transition around $T_S$ is due to on set of the SDW on the Pr sublattice as there is a long-range ordering which sets in at much lower temperature (~5 K).

Phonon anomalies below $T_{MMT}$ ~ 140 K and 130 K for $Pr_2La_2Ni_3O_{10}$ (x = 2) and $Pr_{0.4}La_{3.6}Ni_3O_{10}$ (x = 3.6), respectively; show a deviation from the normal behavior (see Figs. 6(c) and (d). For $Pr_2La_2Ni_3O_{10}$ phonon modes P5, P9, P12 and P14 show strong deviation, as indicated by the shaded region in Fig. 6 (c). For $Pr_{0.4}La_{3.6}Ni_3O_{10}$ phonon modes P4, P9, and P12 exhibit strong deviations, as shown in Fig. 6 (d). In particular, mode P9 for x=2 and x=3.6 demonstrates a relatively small frequency softening rate $\Delta\omega(\omega_{T_{MMT}} - \omega_{6K}) \sim 0.5\,cm^{-1}$ for $T <$ $T_{MMT}$, compared to the softening rate $\Delta\omega(\omega_{330K} - \omega_{T_{MMT}}) \sim 4$ to $5$ cm$^{-1}$ above $T_{MMT}$. We also observe a significant shift in the softening rate for mode P14 for both doping x=2 and x=3.6, the rate remains small $\Delta\omega(\omega_{T_{MMT}} - \omega_{6K}) \sim 2.5$ cm$^{-1}$ for $T < T_{MMT}$, but increases markedly above $T_{MMT}$ i.e., $\Delta\omega(\omega_{330K} - \omega_{T_{MMT}}) \sim 10$ to $14$ cm$^{-1}$ (see Fig. 6(c) and (d)). The significant phonon anomalies observed for all the members of Pr$_{4-x}$La$_x$Ni$_3$O$_{10}$ (x = 0, 0.4, 1, 2, and 3.6) in the metallic states below $T_{MMT}$, interpreted as a couples CDW and SDW, driven by a Fermi-surface nesting picture [17,48].

Using anomalies in the phonon self-energy parameters in the prominent phonon modes for Pr$_{4-x}$La$_x$Ni$_3$O$_{10}$ (x = 0, 1, 2, and 3.6), we extracted the MMT transition temperature labeled as $T_{MMT}$ and the onset of induced SDW ($T_S$) in the Pr sublattice. We find that the MMT transition extracted using Raman data decreases gradually as one moves from x = 0 to x = 3.6, matching quite nicely with the data extracted from the specific heat and resistivity measurement on these samples as described in Fig. 1(d). Figure 7 shows the phase diagram for the evolution of the average metal-to-metal transition temperature and short-range ordering as well as long-range ordering extracted using Raman, specific heat, and resistivity measurements. $T_S$ shows decreasing behaviors as a function of increasing La concentration. Remarkably, For Pr$_{4-x}$La$_x$Ni$_3$O$_{10}$ (x = 0, 0.4, and 1), we find evidence of an additional ordering ($T_S$) well above the respective long-range ordering transition temperature $T_1$, which is reflected in the renormalization of the phonon modes self-energy parameters. As observed from the $C_P/T$ peak at $T_1$, the prominent transition occurs for x = 0 compared to x = 0.4 and x = 1. This is also evident from the significant renormalization observed in the phonon modes of Pr$_4$Ni$_3$O$_{10}$ compared at $T_S$, compared to those in the x = 0.4, and 1 doping samples. At higher temperatures, the SDW exhibits weak correlations and remains quasi-2D. However, as the temperature decreases, the induced order on the Pr sublattice opens new exchange pathways, facilitating the transition to a 3D-ordered state. We note that the variation of softening rate in the intertwined density wave phase and stable metallic phase transition is consistent across all doping concentrations of La in $Pr_{4-x}La_xNi_3O_{10}$ (x = 0, 0.4, 1, 2, 3.6, and 4). Our in-depth analysis reveals a rich phase diagram. In addition to the phonon renormalization around $T_{MMT}$, the effect of induced SDW on Pr sublatticeis is reflected in phonons anamolies observed well above the long-range magnetic ordering temperature. We observed that below $T_{MMT}$, CDW and SDW are intertwined, and below $T_S$ an additional SDW phase becomes active. Our results calls for a thorough theoretical study to analyze how Fermi-surface renormalization below $T_{MMT}$ affects phonon mode softening rate, and we hope our results will motivate such work.

**Conclusion**

In summary, we investigated the lattice dynamics of the trilayer nickelates series $Pr_{4-x}La_xNi_3O_{10}$ (x = 0, 0.4, 1, 2, 3.6, and 4) via lab-based XRD, low temperature specific heat, and temperature dependent polarized Raman scattering. We find the anomalous deviation in the frequency and linewidth of the phonon modes from anharmonic interaction model below the metal-to-metal transition temperature $T_{MMT}$, driven by instabilities in the electronic structure, specifically the Fermi surface, lead to intertwined spin and charge density waves formations. The anomalous linewidth renormalizations below $T_{MMT}$ indicate that the SDW precedes the CDW. The observation of induced SDW ($T_S$) only for the higher doping of Pr (x = 0, 0.4, and 1); far above the long-range ordering ($T_1$) from the anomalous phonon renormalization is attributed to the induced SDW of the Pr sublattice imprinted by the magnetic ordering of Ni sublattice. Additionally, the electronic transition between $4f$ states of the $Pr^{3+}$ ions in the perovskite block layers exhibits anomalous temperature evolution, reflecting an intricate coupling between crystal field excitations and lattice degrees of freedom. Our results unravel the intricate role of CDW and SDW phenomena in $Pr_{4-x}La_xNi_3O_{10}$, and their coupling with the lattice DoF; which offers new insights that we hope will motivate both theoretical and experimental aspects for a further quantitative understanding.

**Acknowledgement:** PK acknowledge financial support from Science and Engineering Research Board (Science and Engineering Research Board - Project no. CRG/2023/002069) and IIT Mandi for the experimental facilities.

**Figures:**

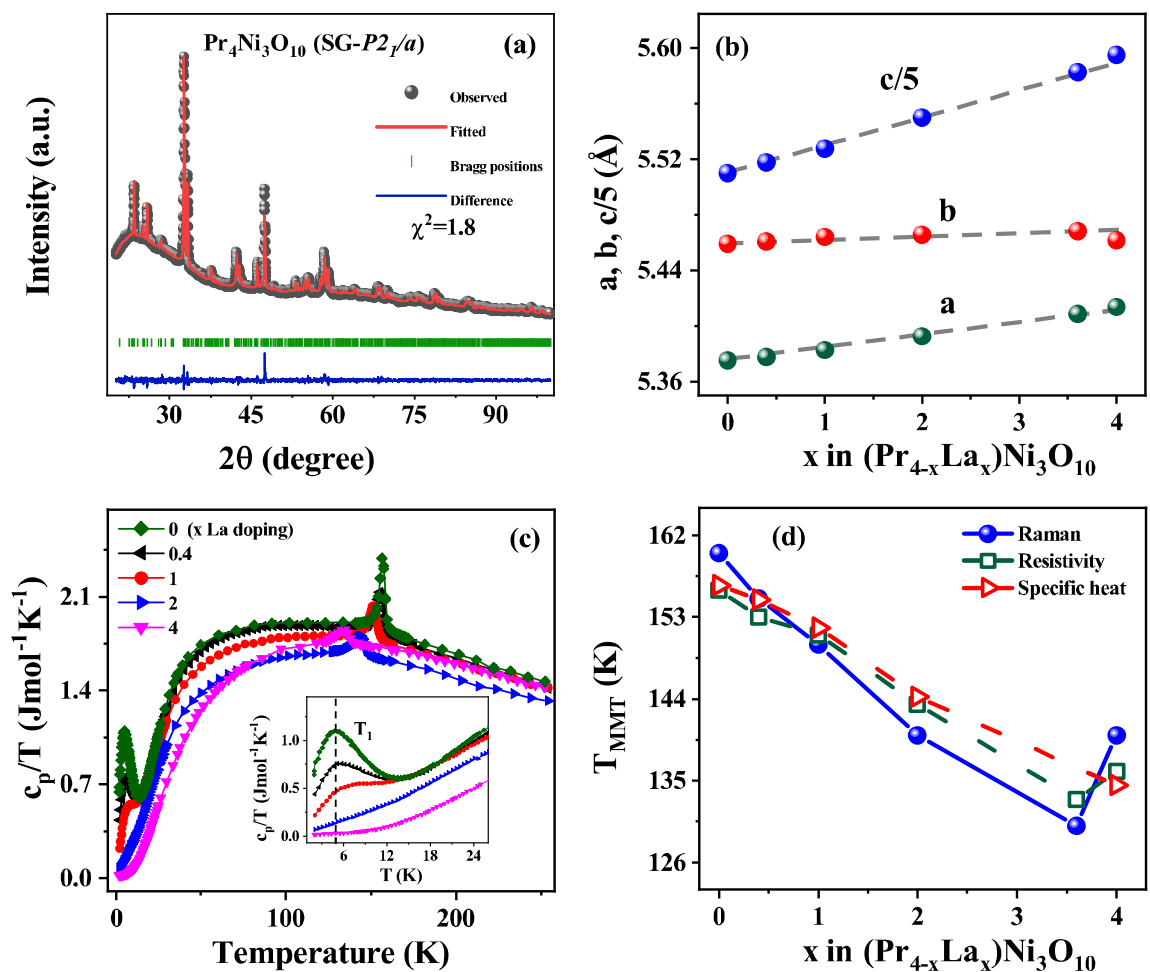

**Figure 1:** (a) Rietveld refinement of the room temperature powder X-ray diffraction (XRD) data of the $Pr_4Ni_3O_{10}$. (b) Lattice parameters (*a*, *b*, and *c*/5) of $Pr_{4-x}La_xNi_3O_{10}$ as a function of doping (x = 0, 0.4, 1, 2, 3.6, and 4). (c) Temperature variation of $c_p/T$ for different La doping. The inset shows the magnification of the low-temperature peak centered at ~ 5 K ($T_1$). (d) Evolution of the $T_{MMT}$ as a function of La doping in $Pr_{4-x}La_xNi_3O_{10}$, extracted using resistivity, specific heat, and Raman measurements.

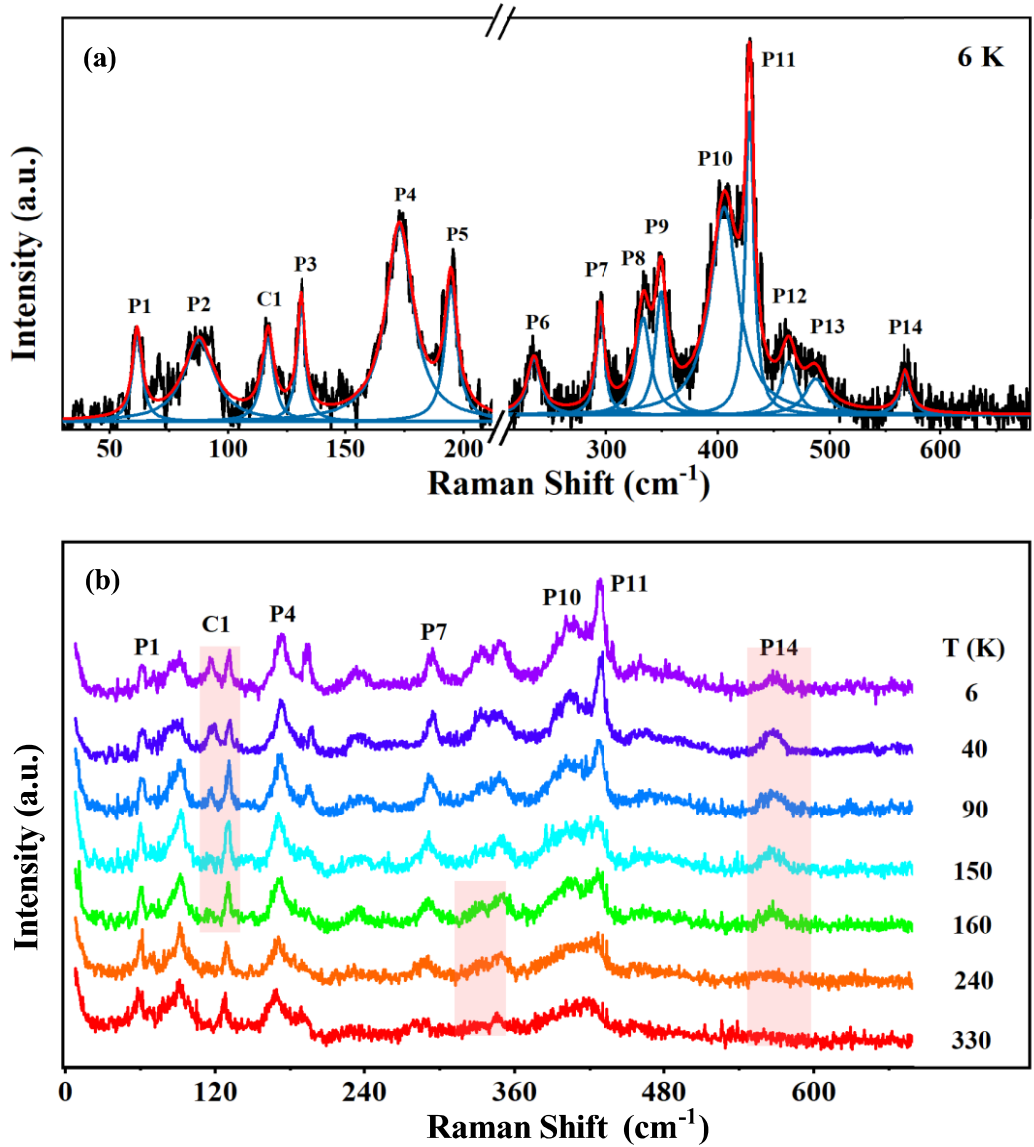

**Figure 2:** (a) Raman spectra of $Pr_4Ni_3O_{10}$ in the spectral range of 30-680 cm$^{-1}$ collected at 6 K. The solid red thick line is a total Lorentzian fit to the experimental data and solid thin blue lines correspond to the individual the phonon modes fits. The observed modes are labeled as P1-P14. (b) Raman spectra of $Pr_4Ni_3O_{10}$ at various temperatures. Red bars mark the phonon modes that gradually disappear with increasing temperature.

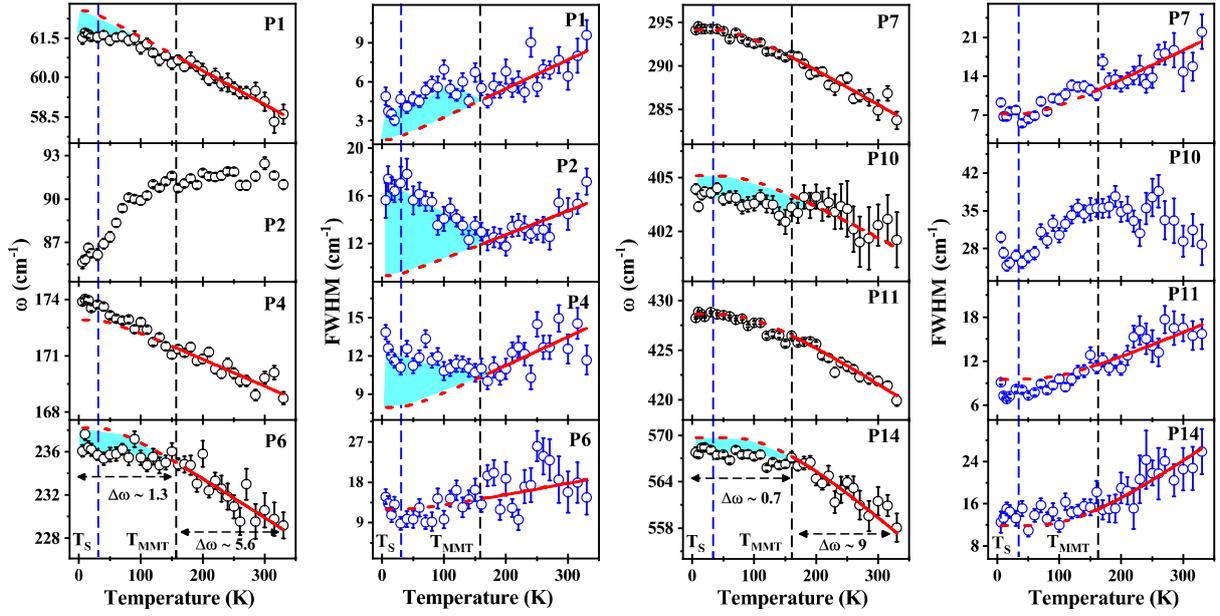

**Figure 3:** Temperature variation of frequencies (ω) and full-width at half maxima (FWHM) for some of the prominent phonon modes of $Pr_4Ni_3O_{10}$. Solid red lines are the fit using the anharmonic model as described in the text and broken red lines are the extrapolation of fitted curves. Error bars are the standard deviation from the Lorentzian fits to the phonon peaks.

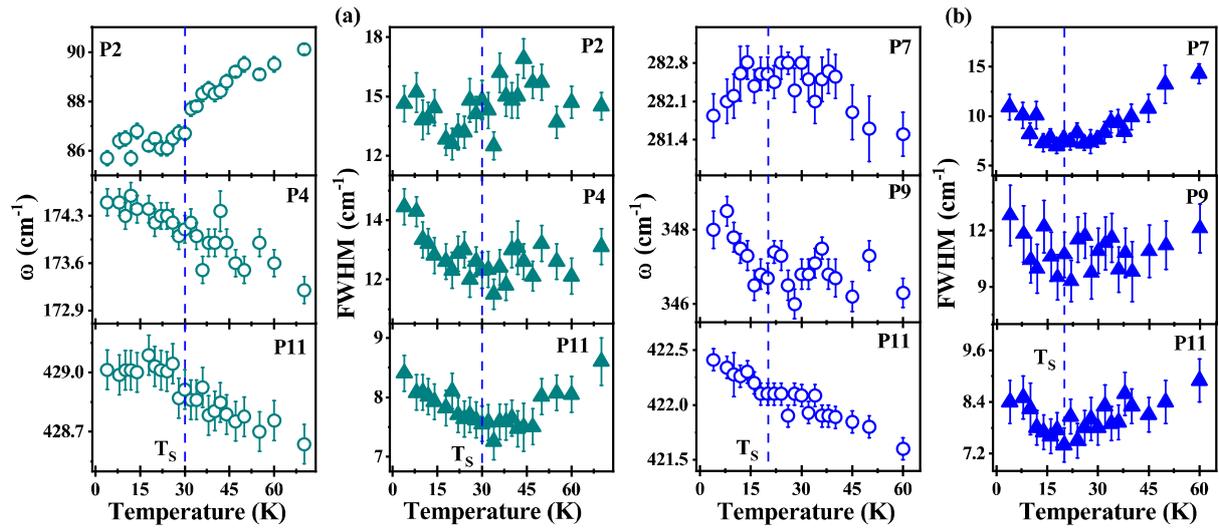

**Figure 4:** Low-temperature (T ≤ 70 K) evolution of frequencies ($\omega$) and full-width at half maxima (FWHM) for some of the prominent phonon modes of (a) $Pr_4Ni_3O_{10}$ (x = 0) (b) $Pr_3La_1Ni_3O_{10}$ (x = 1).

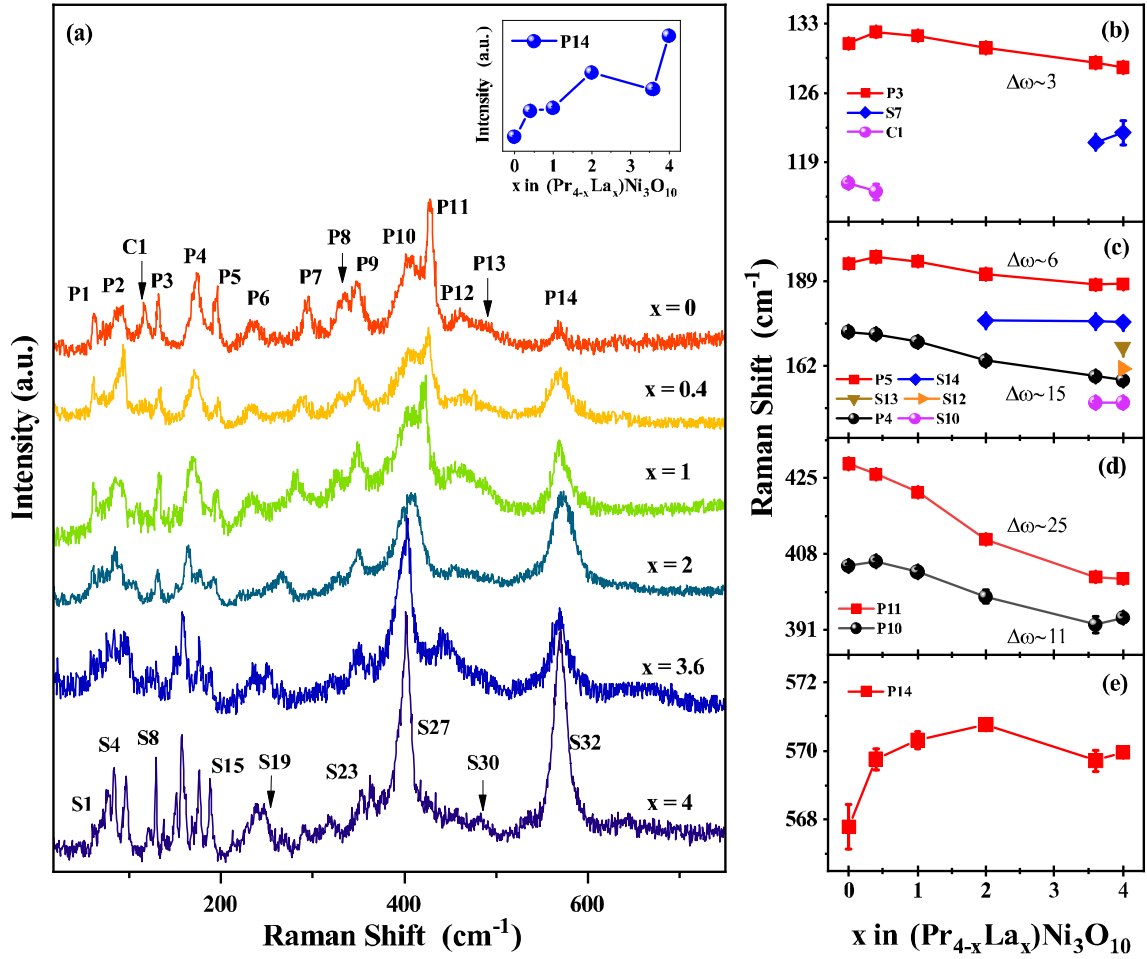

**Figure 5:** (a) Raman spectra of $Pr_{4-x}La_xNi_3O_{10}$ (x = 0, 0.4, 1, 2, 3.6, and 4) at 6 K. Inset shows the intensity evolution of the mode P14 as a function of doping concentration. (b-e) Evolution of the phonon frequencies at 6 K with the doping concentration (see Supplementary Information Fig. S6 for more details on modes labeling).

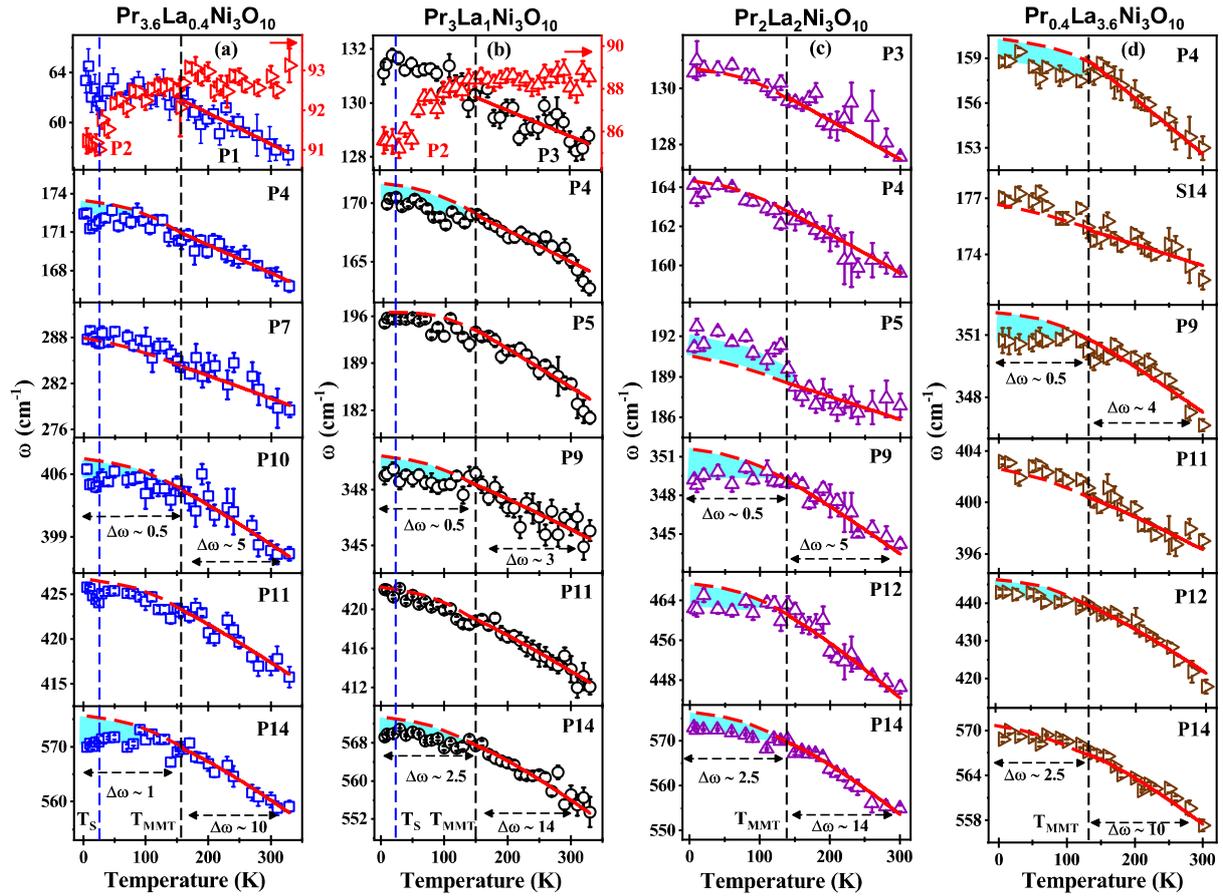

**Figure 6:** (a), (b), (c), and (d) Temperature variation of frequencies ($\omega$) for some of the prominent phonon modes of $Pr_{4-x}La_xNi_3O_{10}$ for x = 0.4, 1, 2, and 3.6, respectively. Solid red lines are fit to the anharmonic model as described in the text, and broken red lines are the extrapolated fitted curves. Error bars are the standard deviation from the Lorentzian fits to the phonon peaks.

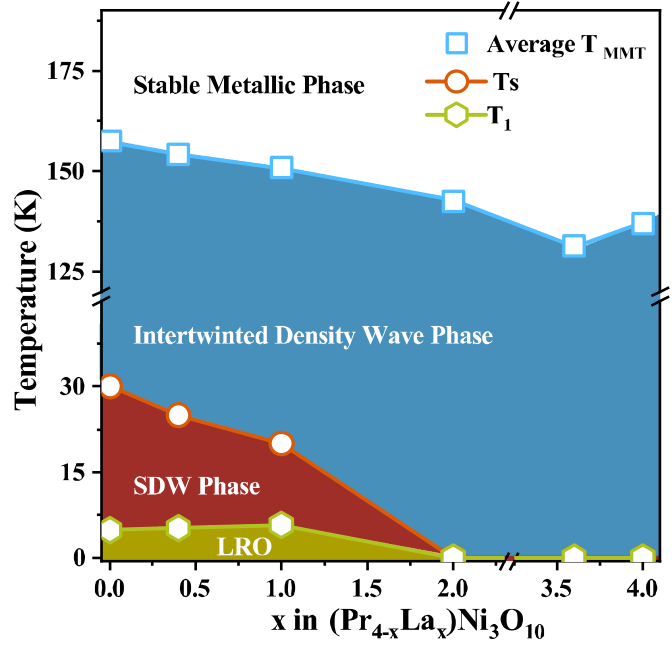

**Figure 7:** Evolution of the average metal-to-metal transition ($T_{MMT}$) temperature extracted from specific heat, resistivity, and Raman measurements. The blue region represents the intertwined phase. The red-shaded region denotes the spin-density wave phase on the Pr sublattice, along with the evolution of the SDW order transition ($T_S$) extracted using Raman measurement, denoted by the red circle. The green shaded region indicates the long-range ordering phase below the temperature $T_1$ denoted by the green hexagonal symbol.

# Supplementary Information:

# Interplay of phonons, intertwined density waves, and induced spin density wave in trilayer nickelates $Pr_{4-x}La_xNi_3O_{10}$

Sonia Deswal[1, †], Dibyata Rout[2], Nirmalya Jana[3], Koushik Pal[3], Surjeet Singh[2] and Pradeep Kumar[1, *]

[1] *School of Physical Sciences, Indian Institute of Technology Mandi, Mandi-175005, India*
[2] *Department of Physics, Indian Institute of Science Education and Research Pune, Pune-411008, India*
[3] *Department of Physics, Indian Institute of Technology Kanpur, Kanpur-208016, India*

†soniadeswal255@gmail.com
* pkumar@iitmandi.ac.in

## S1. Polarization-dependent measurement

To investigate the symmetry of the lattice vibrations, we performed polarization-dependent Raman measurements of the undoped (x = 0) $Pr_4Ni_3O_{10}$ system. In the scattering configuration $(\theta\ \theta_0)$, the incident light polarization is rotating at an angle $(\theta)$ using a polarizer ($\lambda/2$ plate) in steps of 20° from 0° to 360°; while the position of the sample and direction of scattered light's polarization is fixed parallel to the X using an analyzer. According to the semiclassical approach the Raman scattering cross section is given as $I_{Raman} \propto \left|\hat{e}_s^T . R . \hat{e}_i\right|^2$, where '*T*' symbolizes transpose, $\hat{e}_i$ and $\hat{e}_s$ are the polarization vectors of the incident and scattered light, respectively, and '*R*' is the Raman tensor [1-3]. In the matrix form, the polarization unit vectors are given as: $\hat{e}_i = \left[\cos(\theta_0 + \theta)\ \sin(\theta_0 + \theta)\ 0\right]$; $\hat{e}_s = \left[\cos(\theta_0)\ \sin(\theta_0)\ 0\right]$, where $\theta$ is the relative angle between $\hat{e}_i$ and $\hat{e}_s$, and $\theta_0$ is an arbitrary angle of scattered light from the X-axis as per the schematic shown in Fig. S3(a). The Raman tensors for the phonon modes are listed in Table S1. Using the above expression angular dependence of the Raman intensity for $A_g$ and $B_g$ modes are given as:

$$I_{A_g} = \left\|\left[(a\cos\theta_0 + d\sin(\theta_0))\cos(\theta_0 + \theta) + (d\cos\theta_0 + b\sin(\theta_0))\sin(\theta_0 + \theta)\right]\right\|^2$$
$$\equiv \left\|[a\cos(\theta) + d\sin(\theta)]\right\|^2$$

$$I_{B_g} = 0$$

Without the loss of generality $\theta_0$ maybe taken to as zero, giving rise to the expression for Raman intensity as $I_{A_g} = \left\|[a\cos(\theta) + d\sin(\theta)]\right\|^2$. Figure S3(b), shows the polarized Raman spectra for different configurations at 6 K and shows that all phonon modes intensity reduces to nearly zero in $\bar{Z}(XY)Z$ configuration. The intensity of $A_g$ modes show a twofold symmetric nature, i.e. it has a maximum intensity at both 0° and 180°, while the intensity approaches zero at 90° and 270°. Figure S3(c), shows the 2D color contour maps of the Raman intensity vs Raman shift as a function of polarization angle. The polarization-dependent results attribute that all modes follow the $A_g$ symmetry.

**Figures:**

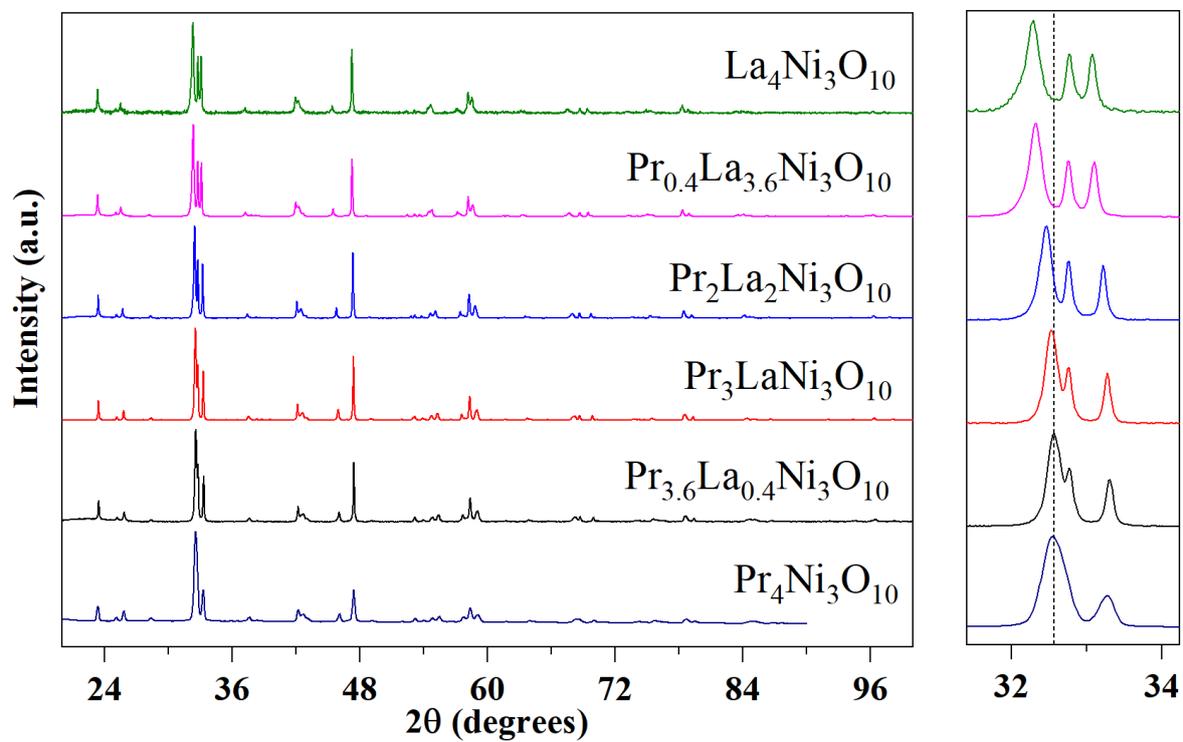

**Figure S1:** Room temperature XRD data for $Pr_{4-x}La_xNi_3O_{10}$ (x = 0, 0.4, 1, 2, 3.6, and 4).

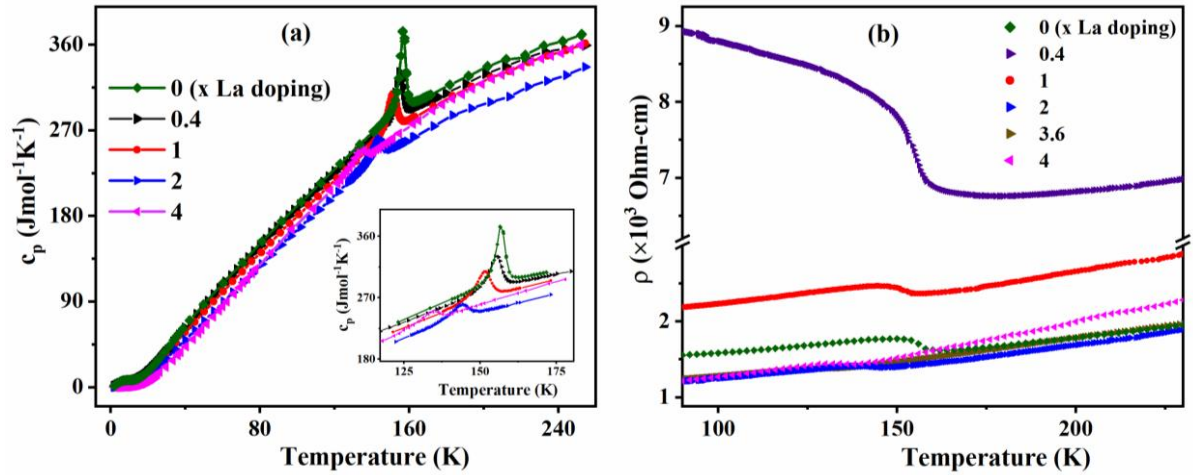

**Figure S2:** (a) and (b) Temperature variation of specific heat ($c_p$) and resistivity for different La doping of $Pr_{4-x}La_xNi_3O_{10}$, respectively.

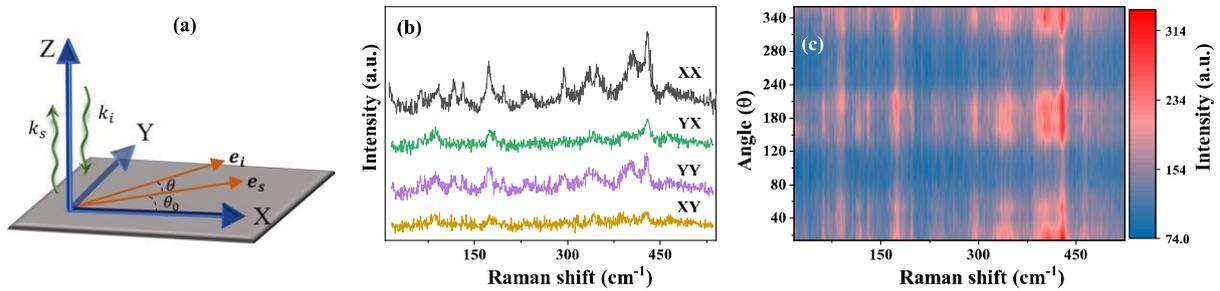

**Figure S3:** (a) The plane projection of the scattered and the incident light polarization direction. (b) Raman spectra of $Pr_4Ni_3O_{10}$ at 6 K taken under different polarization configurations. (c) The two-dimensional color contour maps of the Raman intensity vs Raman shift as a function of the polarization angle.

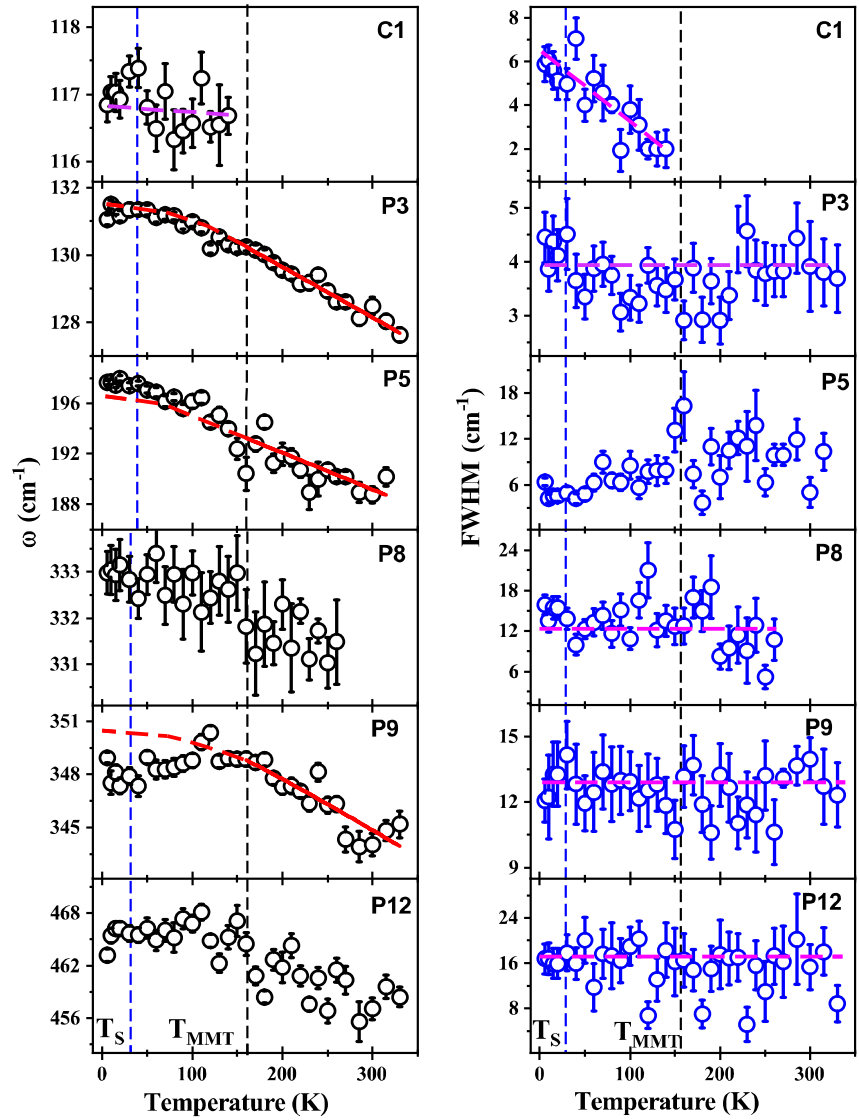

**Figure S4:** Temperature variation of the frequencies ($\omega$) and full-width at half maxima (FWHM), for some of the phonon modes of undoped (x = 0) $Pr_4Ni_3O_{10}$. Solid red lines are fit to the anharmonic model described in the text, and broken red lines are the extrapolated fitted curves till low temperature. Error bars are the standard deviation from the Lorentzian fits to the phonon peaks. The broken pink lines are drawn as guide to the eye.

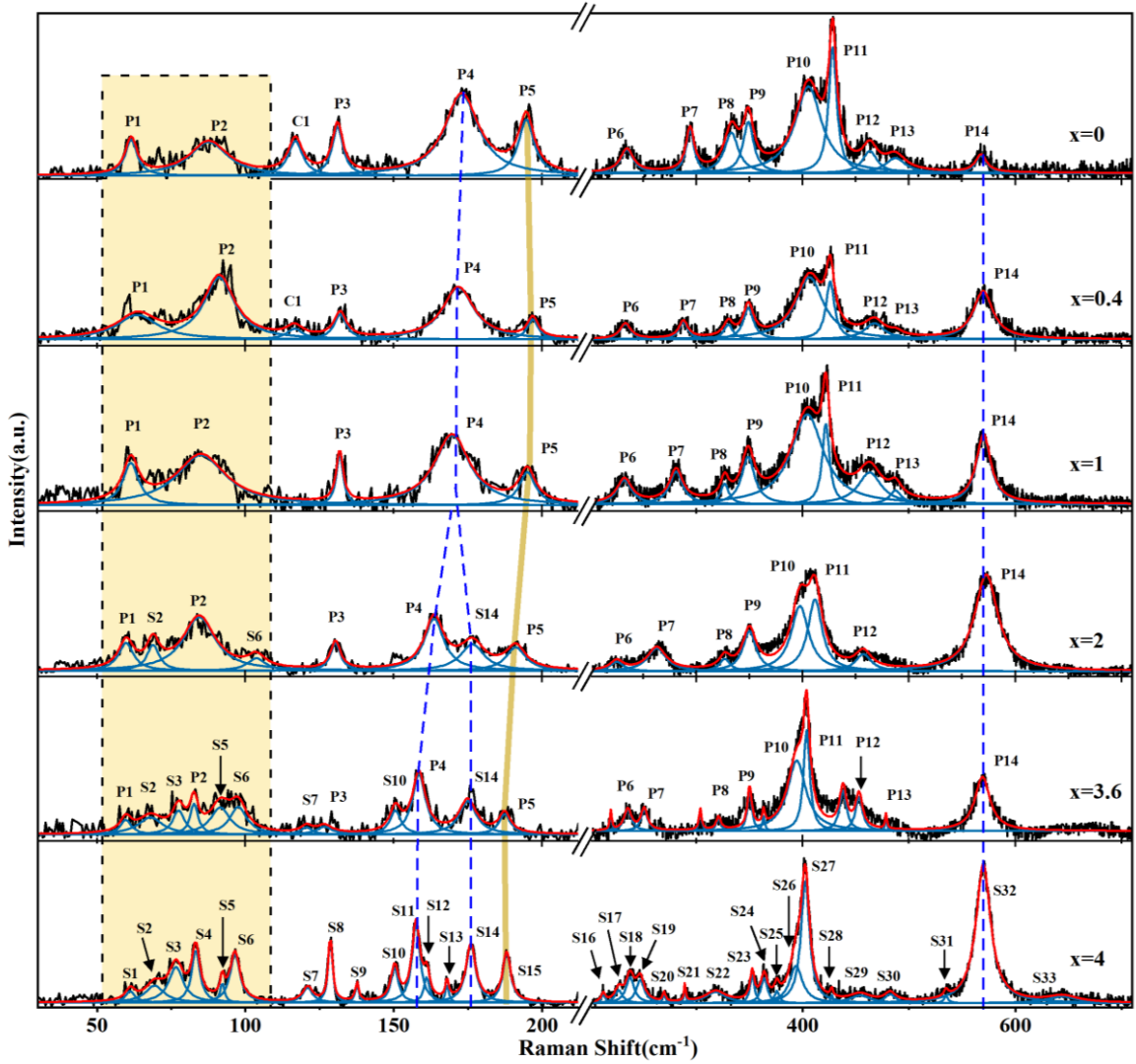

**Figure S5:** Raman spectra of $Pr_{4-x}La_xNi_3O_{10}$ (x = 0, 0.4, 1, 2, 3.6, and 4) in 30-700 cm$^{-1}$ spectral range collected at 6 K. The solid red thick line is a total sum of the Lorentzian fit to the experimental data and solid thin blue lines correspond to the individual fit of the phonon modes. The observed modes for (x = 0) and (x = 4) are labeled as P1-P14 and S1-S33, respectively.

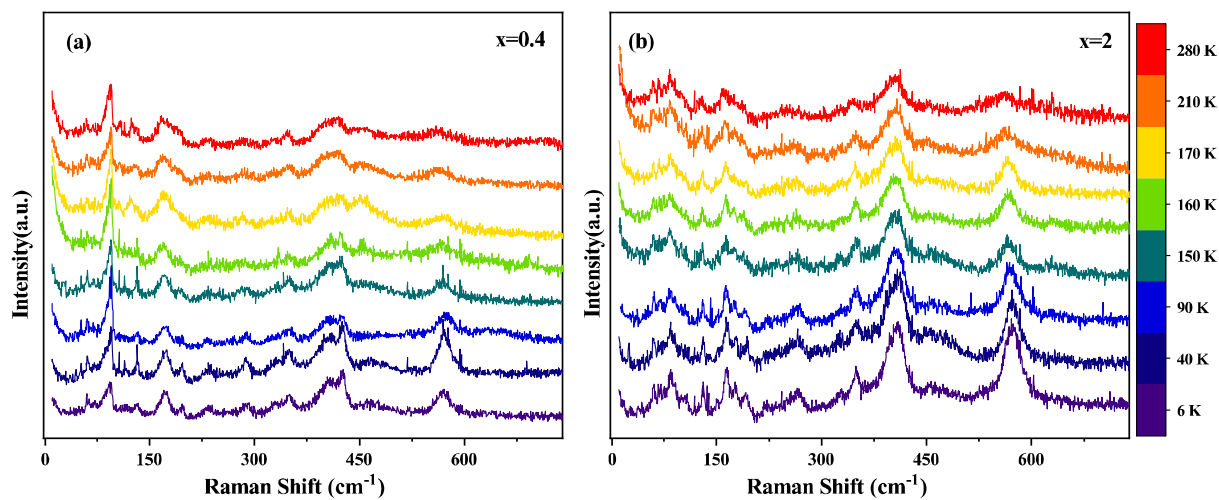

**Figure S6:** Temperature evolution of the Raman spectra of $Pr_{4-x}La_xNi_3O_{10}$ (x = 0.4, 2) in the spectral range of 10-740 cm$^{-1}$.

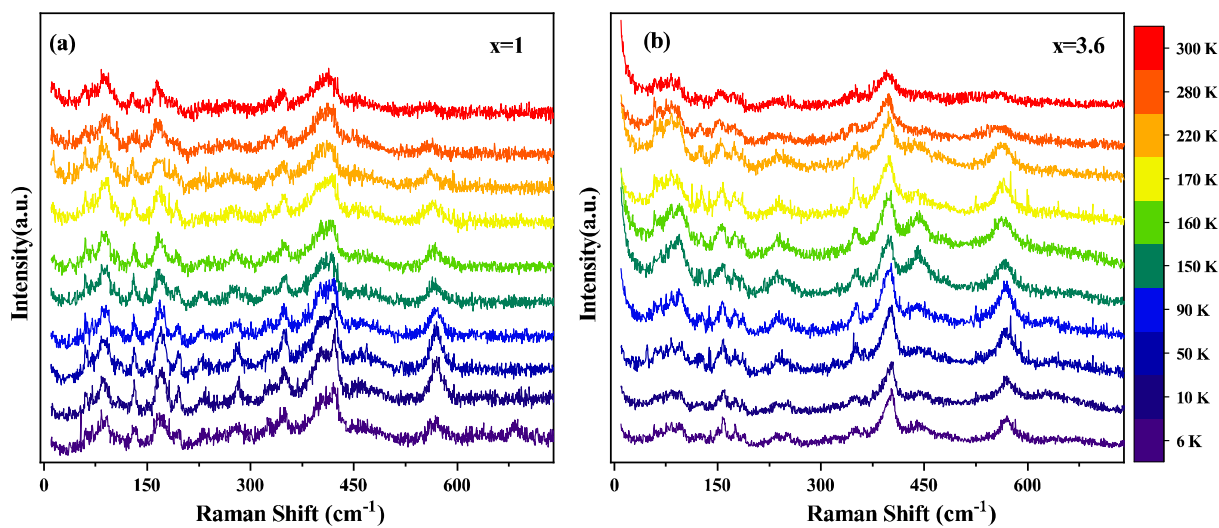

**Figure S7:** Temperature evolution of the Raman spectra of $Pr_{4-x}La_xNi_3O_{10}$ (x = 1, 3.6) in the spectral range of 10-740 cm$^{-1}$.

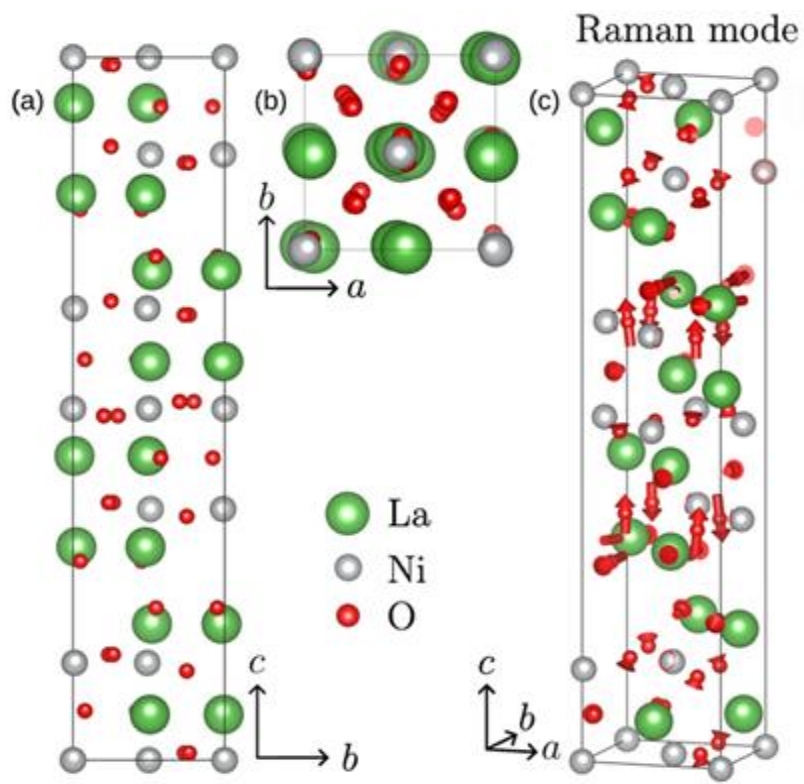

**Figure S8:** (a) The side view of the unit cell of $La_4Ni_3O_{10}$, (b) The top view of the unit cell. (c) The low-frequency (30.7 cm$^{-1}$) Raman-active mode, the vectors highlight the directions of atomic displacement and their magnitude.

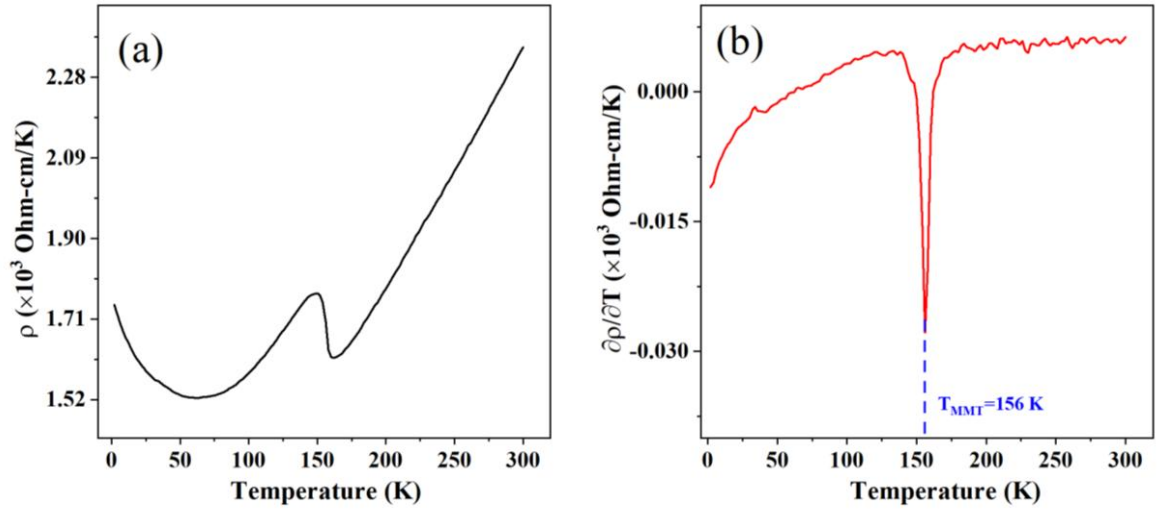

**Figure S9**: (a) Temperature dependence of resistivity for $Pr_4Ni_3O_{10}$. (b) Temperature dependence of ($d\rho/dT$), a clear drop shows the onset temperature $T_{MMT}$.

**Tables:**

**Table S1:** Atoms and the corresponding Wyckoff positions in the unit cell and irreducible representations of the phonon modes at $\Gamma$ point for the monoclinic (space group $P2_1/a$) trilayernickelate $Pr_4Ni_3O_{10}$. Irreducible representation $\Gamma_{Total}, \Gamma_{Raman}, \Gamma_{IR}$ and $\Gamma_{Acoustic}$ corresponds to total, Raman, Infrared, and acoustic phonon modes representation at the $\Gamma$ point, respectively. In the first coloum, second digit in the bracket represent number of that particular atom in the unit cell.

| Monoclinic *P2₁/a* (14) | | | |
|---|---|---|---|
| **Atom** | **Wyckoff site** | **Point mode Decomposition** | **Raman Tensors** |
| **Pr[1,4]** | 4e | $3A_g + 3A_u + 3B_g + 3B_u$ | $R_{A_g} = \begin{pmatrix} a & d & 0 \\ d & b & 0 \\ 0 & 0 & c \end{pmatrix}$ |
| **Ni(1)** | 2a | $3A_u + 3B_u$ | |
| **Ni(2)** | 2b | $3A_u + 3B_u$ | |
| **Ni(3,4)** | 4e | $3A_g + 3A_u + 3B_g + 3B_u$ | $R_{B_g} = \begin{pmatrix} 0 & 0 & e \\ 0 & 0 & f \\ e & f & 0 \end{pmatrix}$ |
| **O[1,10]** | 4e | $3A_g + 3A_u + 3B_g + 3B_u$ | |

$\Gamma_{Total} = 48A_g + 54A_u + 48B_g + 54B_u$, $\Gamma_{Raman} = 48A_g + 48B_g$, $\Gamma_{IR} = 53A_u + 52B_u$, $\Gamma_{Acoustical} = A_u + 2B_u$

**Table S2:** List of the experimentally observed modes for La$_4$Ni$_3$O$_{10}$ at 6K, along with DFT bsaed calculated zone centered mode frequencies. Unit in cm$^{-1}$.

| Mode Assignment | Experimentally $\omega$ ($cm^{-1}$) (6 K) | DFT $\omega$ ($cm^{-1}$) | Mode Assignment | Experimentally $\omega$ ($cm^{-1}$) (6 K) | DFT $\omega$ ($cm^{-1}$) |
|---|---|---|---|---|---|
| S1 | 61.5±0.4 | - | S18 | 237.7±0.4 | 237.5 |
| S2 | 69.5±0.6 | 69.4 | S19 | 247.3±0.3 | 246 |
| S3 | 76.7±0.2 | 77.9 | S20 | 269.7±0.7 | 271.5 |
| S4 | 83.3±0.09 | 88.3 | S21 | 290.0±0.5 | 292.4 |
| S5 | 92.2±0.2 | 90.9 | S22 | 319.2±0.7 | 319.2 |
| S6 | 96.5±0.1 | 97.9 | S23 | 352.5±0.4 | 352.8 |
| S7 | 121.0±0.3 | 121.5 | S24 | 364.4±0.3 | 361.6 |
| S8 | 128.6±0.04 | 127.1 | S25 | 375.8±0.1 | 375.2 |
| S9 | 137.6±0.1 | 138.4 | S26 | 393.9±0.6 | 394.2 |
| S10 | 150.3±0.1 | 150.3 | S27 | 402.4±0.1 | 403.4 |
| S11 | 157.5±0.1 | 157.9 | S28 | 429.7±0.5 | 432.1 |
| S12 | 161.1±0.1 | 161.6 | S29 | 454.7±1.0 | 458.9 |
| S13 | 168.0±0.1 | 166.6 | S30 | 481.9±0.9 | 480.5 |
| S14 | 175.9±0.1 | 175.1 | S31 | 535.0±0.9 | 523.9 |
| S15 | 188.0±0.1 | 189.7 | S32 | 569.9±0.07 | 568.5 |
| S16 | 213.4±1.0 | 209.8 | S33 | 643.9±1.4 | - |
| S17 | 227.3±0.5 | 226.8 | | | |

**Table S3:** List of the experimentally observed modes with their frequencies ($\omega$) at 6K for Pr$_{4-x}$La$_x$Ni$_3$O$_{10}$ (x = 0, 0.4, 1, 2, 3.6, and 4). Unit in cm$^{-1}$. The highlighted modes represent the prominent phonon modes that remain consistent across all doping levels.

| Mode Assignment | $\omega$ (cm$^{-1}$) x = 4 La$_4$Ni$_3$O$_{10}$ | $\omega$ (cm$^{-1}$) x = 3.6 | $\omega$ (cm$^{-1}$) x = 2 | $\omega$ (cm$^{-1}$) x = 1 | $\omega$ (cm$^{-1}$) x = 0.4 | $\omega$ (cm$^{-1}$) x = 0 Pr$_4$Ni$_3$O$_{10}$ | Mode Assignment |
|---|---|---|---|---|---|---|---|
| S1 | 61.5±0.4 | 59.8±0.6 | 59.7±0.4 | 61.5±0.3 | 63.2±0.7 | 61.5±0.2 | **P1** |
| S2 | 69.5±0.6 | 67.2±0.4 | 68.6±0.5 | | | | |
| S3 | 76.7±0.2 | 77±0.6 | | | | | |
| S4 | 83.3±0.09 | 82.8±0.2 | 84.5±0.3 | 85.4±0.5 | 91.2±0.2 | 87.6±0.4 | **P2** |
| S5 | 92.2±0.2 | 90.8±1.1 | | | | | |
| S6 | 96.5±0.1 | 98±0.7 | 104.4±0.6 | | | | |
| | | | | | 116±0.8 | 116.9±0.2 | **C1** |
| S7 | 121.0±0.3 | 121.9±1.2 | | | | | |
| S8 | 128.4±0.04 | 129±0.2 | 130.6±0.2 | 131.8±0.2 | 132.1±0.4 | 131.0±0.2 | **P3** |
| S9 | 137.6±0.1 | | | | | | |
| S10 | 150.3±0.1 | 150.2±0.4 | | | | | |
| S11 | 157.5±0.1 | 158.7±0.2 | 163.6±0.2 | 169.6±0.3 | 172.0±0.3 | 172.9±0.2 | **P4** |
| S12 | 161.1±0.1 | | | | | | |
| S13 | 168.0±0.1 | | | | | | |
| S14 | 175.9±0.1 | 176.1±0.2 | 176.4±0.4 | | | | |
| S15 | 188.0±0.1 | 187.8±0.4 | 191.2±0.4 | 195.3±0.3 | 196.8±0.4 | 194.7±0.2 | **P5** |
| S16 | 213.4±1.0 | 219.2±0.8 | | | | | |
| S17 | 227.3±0.5 | | | | | | |
| S18 | 237.7±0.4 | 235.1±0.7 | 238.6±3.3 | 232.5±0.6 | 233.9±0.7 | 234.7±0.5 | **P6** |
| S19 | 247.3±0.3 | 251.4±0.5 | 264.8±0.5 | 280.8±0.5 | 287.7±0.6 | 294.1±0.2 | **P7** |
| S20 | 269.7±0.7 | | | | | | |
| S21 | 290.0±0.5 | 302.9±3.7 | | | | | |
| S22 | 319.2±0.7 | 320.6±1.1 | 326.7±0.7 | 327.4±0.7 | 329.5±0.6 | 332.9±0.5 | **P8** |
| S23 | 352.5±0.4 | 349.3±0.6 | 349.2±1.3 | 348.8±0.4 | 348.7±0.4 | 348.9±0.3 | **P9** |
| S24 | 364.4±0.3 | | | | | | |
| S25 | 375.8±0.1 | 363.3±0.4 | | | | | |
| S26 | 393.9±0.6 | 392.2±1.8 | 398.5±1.5 | 404.0±0.6 | 406.4±0.5 | 405.4±0.3 | **P10** |
| S27 | 402.4±0.1 | 402.8±0.3 | 411.2±0.6 | 421.9±0.2 | 425.8±0.2 | 428.2±0.1 | **P11** |
| S28 | 429.7±0.5 | 437.4±8.6 | | | | | |
| S29 | 454.7±1.0 | 449.9±9.3 | 460.3±1.4 | 463.7±1.0 | 467.4±0.9 | 463.2±0.7 | **P12** |
| S30 | 481.9±0.9 | 481.5±1.4 | | 487.7±1.4 | 487.9±0.4 | 487.2±1.4 | **P13** |
| S31 | 535.0±0.9 | | | | | | |
| S32 | 569.9±0.07 | 569.7±0.3 | 570.8±0.2 | 570.3±0.2 | 569.8±0.3 | 567.8±0.7 | **P14** |
| S33 | 643.9±1.4 | | | | | | |